\documentclass[prd,preprint,nofootinbib]{revtex4}

\usepackage{graphicx}
\usepackage{amssymb}
\usepackage{amsmath}
\usepackage{color}

\IfFileExists{srcltx.sty}{\usepackage[active]{srcltx}}

\begin{document}
\renewcommand{\thefootnote}{\#\arabic{footnote}}
\newcommand{\rem}[1]{{\bf [#1]}}
\newcommand{\gsim}{ \mathop{}_ {\textstyle \sim}^{\textstyle >} }
\newcommand{\lsim}{ \mathop{}_ {\textstyle \sim}^{\textstyle <} }
\newcommand{\vev}[1]{ \left\langle {#1}  \right\rangle }
\newcommand{\bear}{\begin{array}}  
\newcommand {\eear}{\end{array}}
\newcommand{\bea}{\begin{eqnarray}}   
\newcommand{\eea}{\end{eqnarray}}
\newcommand{\beq}{\begin{equation}}   
\newcommand{\eeq}{\end{equation}}
\newcommand{\bef}{\begin{figure}}  
\newcommand {\eef}{\end{figure}}
\newcommand{\bec}{\begin{center}} 
\newcommand {\eec}{\end{center}}
\newcommand{\non}{\nonumber}  
\newcommand {\eqn}[1]{\beq {#1}\eeq}
\newcommand{\la}{\left\langle}  
\newcommand{\ra}{\right\rangle}
\newcommand{\ds}{\displaystyle}
\newcommand{\red}{\textcolor{red}}
\def\SEC#1{Sec.~\ref{#1}}
\def\FIG#1{Fig.~\ref{#1}}
\def\EQ#1{Eq.~(\ref{#1})}
\def\EQS#1{Eqs.~(\ref{#1})}
\def\lrf#1#2{ \left(\frac{#1}{#2}\right)}
\def\lrfp#1#2#3{ \left(\frac{#1}{#2} \right)^{#3}}
\def\GEV#1{10^{#1}{\rm\,GeV}}
\def\MEV#1{10^{#1}{\rm\,MeV}}
\def\KEV#1{10^{#1}{\rm\,keV}}
\def\REF#1{(\ref{#1})}
\def\lrf#1#2{ \left(\frac{#1}{#2}\right)}
\def\lrfp#1#2#3{ \left(\frac{#1}{#2} \right)^{#3}}
\def\OG#1{ {\cal O}(#1){\rm\,GeV}}

\begin{flushright}
IPMU 10-0095
\end{flushright}

\title{
Dark Matter from Split Seesaw
}

\author{Alexander Kusenko$^{(a,b)}$, Fuminobu Takahashi$^{(b)}$ and  Tsutomu T. Yanagida$^{(b,c)}$}

\affiliation{
${}^{(a)}$Department of Physics \& Astronomy, University of California, Los
 Angeles, CA 90095, USA\\
${}^{(b)}$Institute for the Physics and Mathematics of the Universe,
University of Tokyo, Chiba 277-8583, Japan\\
${}^{(c)}$Department of Physics,
University of Tokyo, Tokyo 113-0033, Japan
}

\begin{abstract}

The seesaw mechanism in models with extra dimensions is shown to be
generically consistent with a broad range of Majorana masses.  The resulting 
democracy of scales implies that the seesaw mechanism can naturally explain the smallness 
of neutrino masses for an arbitrarily small right-handed neutrino mass. If the
scales of the seesaw parameters are split, with two right-handed 
neutrinos at a high scale and one at a keV scale, one can explain the
matter-antimatter asymmetry of the universe, as well as dark matter.
The dark matter candidate, a sterile right-handed neutrino with mass of several
keV, can account for the observed pulsar velocities and for the recent data
from {\em Chandra} X-ray Observatory, which suggest the existence of a 5~keV  
sterile right-handed neutrino.

\end{abstract}


\maketitle

\section{Introduction}
\label{sec:1}

The possible existence of extra dimensions has changed the way one
thinks about the hierarchy of scales in particle
physics~\cite{ArkaniHamed:1998rs,Randall:1999ee}.  If the light
Standard Model (SM) degrees of freedom are localized on a
(3+1)-dimensional brane embedded in a higher-dimensional spacetime,
while any other degrees of freedom reside on other branes or propagate
in the bulk, the perceived difference between the high scale and the
low scale may arise from the suppression that is exponential in the
size of the extra dimension.  It is a generic phenomenon in this class
of models that the interactions of fields localized on different
branes can be exponentially suppressed by the distance between the
branes~\cite{Hebecker:2002xw}.  In the low-energy effective theory, the
smallness of the couplings may seem completely unnatural, because the
symmetry group of the (3+1)-dimensional effective theory need not
increase in the limit where the size of the extra dimension becomes
large.

Neutrino masses and the possibility of leptogenesis offer an
intriguing connection to the high-scale physics.  In (3+1) dimensions,
the naturalness arguments suggest that the SU(2) singlet fermions
should have masses of the order of the high scale, in contrast with
the non-singlet fermions, whose masses are protected by the chiral
gauge SU(2) symmetry. However, if extra dimensions exist, one must
rethink the naturalness arguments. A fermion localized on a distant
brane may appear as weakly interacting light particle in the
low-energy effective four-dimensional Lagrangian.

Let us review the natural values of parameters in the modern version
of the Standard Model, by which we mean the original Standard Model of
Glashow, Salam, and Weinberg, supplied with three right-handed
neutrinos generating the observed neutrino masses via the seesaw
mechanism~\cite{seesaw}.  As mentioned above, and as we will show in
detail below, both heavy and light right-handed neutrinos can be
equally natural if extra dimensions exist, assuming that the
small-mass fermions in (3+1) dimensions arise from the heavy particles
on a remote brane.  The simultaneous existence of light and heavy
right-handed neutrinos allows one to explain both the baryon asymmetry
of the universe and dark matter in the framework of the simplest model
consistent with the data, -- the modern variant of the Standard Model.
The dark matter candidate in this model is a keV sterile neutrino, the
possible existence of which is supported by some astrophysical
data~\cite{Kusenko:2009up,Kusenko:1997sp,reion,Loewenstein:2009cm}.

The only way to generate the baryon asymmetry of the universe in this
minimal scenario is via leptogenesis~\cite{Fukugita:1986hr}, because
the lack of CP violation and the weakness of the electroweak phase
transition eliminate any alternative scenarios.  Only two right-handed
neutrinos are necessary for leptogenesis~\cite{Frampton:2002qc}.
However, having three right-handed neutrinos, one for each generation
of fermions, is arguably more appealing; this also allows one to embed
the Standard Model into SO(10) Grand Unified Theory or any theory that
has a gauge $U(1)_{\rm B-L}$ symmetry.  What is the natural value for
the Majorana mass of the third right-handed field? In the presence of
extra dimensions, the high-scale values are not necessarily favored by
any naturalness arguments.  However, if this Majorana mass is of the
order of keV, one can explain dark matter, as well as the
long-standing puzzle of the pulsar kicks~\cite{Kusenko:1997sp,Kusenko:2009up}.  
The same particles can play an important role in the formation of the first stars~\cite{reion} 
and other astrophysical phenomena~\cite{Biermann:BH}.
Furthermore, at least two
recent astrophysical observations suggest the existence of a sterile
neutrino either with 5 keV mass~\cite{Loewenstein:2009cm} or 17 keV
mass~\cite{Prokhorov:2010us}.

By splitting the scales of the heavy and light right-handed neutrinos,
which is naturally realized in the {\em split seesaw} model described below, 
we achieve an elegant and simple description of all known experimental data. 
We note that models with three right-handed neutrinos below the electroweak scale have 
been proposed~\cite{deGouvea:2005er,Asaka:2005an}, and in one of them, 
dubbed $\nu$MSM~\cite{Asaka:2005an}, one can explain the baryon asymmetry by demanding a
high degree of degeneracy between two Majorana masses, both of the
order of several GeV.  The high degree of degeneracy is required to
amplify the effects of CP violation in a leptogenesis scenario that
involves oscillations~\cite{Akhmedov:1998qx}.  In contrast, we employ
the standard leptogenesis using decays of the heavy right-handed
neutrinos, and no mass degeneracy is required. We emphasize, however, that 
our model with an extra dimension is not limited to the conventional leptogenesis scenario, 
and it can explain a very broad range of scales. 
Indeed, our model assures that the beauty of the seesaw formula, which explains the smallness of the neutrino masses 
by relating them to the ratio of the weak scale and the GUT scale, is preserved for practically arbitrary choice of 
the right-handed neutrino masses.

\section{Split seesaw mechanism}
\label{sec:2}
We begin by reviewing the seesaw mechanism in the 4D theory. 
The relevant terms in the Lagrangian are given by
\beq
{\cal L} \;=\; 
i {\bar N}_i  \gamma^\mu \partial_\mu N_i+
\left(
 \lambda_{i \alpha} {\bar N}_i L_\alpha \, \phi 
-\frac{1}{2} M_{Ri} {\bar {N^c_i}} N_i + {\rm h.c.}
\right),
\label{4dL}
\eeq
where $N_i$, $L_\alpha$ and $\phi$ are the right-handed neutrino,
lepton doublet and Higgs boson, respectively, $i$ denotes the
generation of the right-handed neutrino, and $\alpha$ runs over the
lepton flavor, $e$, $\mu$ and $\tau$.  Integrating out the massive
right-handed neutrinos yields the seesaw formula for the light
neutrino mass:
\bea
\left(m_\nu\right)_{\alpha \beta} &=& \sum_i \lambda_{i\alpha}\lambda_{i\beta} \frac{\la\phi^0\ra^2 }{M_{Ri}}.
\label{seesaw}
\eea
The atmospheric~\cite{Hosaka:2006zd} and solar~\cite{Hosaka:2005um,Aharmim:2005gt} neutrino oscillation experiments have
provided firm evidence that at least two neutrinos have small but
non-zero masses, and the mass splittings are given by $\Delta m^2_{\rm
  atm} \simeq 2 \times 10^{-3} {\rm eV}^2$ and $\Delta m^2_\odot
\simeq 8 \times 10^{-5} {\rm eV}^2$.  The seesaw mechanism then
suggests that a typical mass scale of the right-handed neutrinos should be $\sim
10^{15}$\,GeV, close to the GUT scale, for $\lambda_{i \alpha} \sim 1$.

To explain the neutrino oscillation data, one must 
introduce more than one right-handed neutrino.  It was shown in
Ref.~\cite{Frampton:2002qc} that two right-handed neutrinos, $N_2$ and
$N_3$, would suffice for this purpose.  Moreover, it is possible to
generate the cosmological baryon asymmetry via leptogenesis with the
two right-handed neutrinos.  Thus, the addition of the two
right-handed neutrinos appears to be the most economical extension of the SM.

However there are two issues in the above model, which, as we will
see below, can be addressed simultaneously and naturally in a theory with an 
extra dimension.  One issue is that at least the lightest of $N_{2,3}$ must 
have a mass $O(10^{11-12})$\,GeV, several orders of
magnitude below the GUT scale, for the leptogenesis to
work~\cite{Endoh:2002wm,Raidal:2002xf}.  Therefore, to explain the neutrino masses, 
one must suppress both the right-handed Majorana mass and the associated Yukawa couplings.  
This is not a severe fine-tuning, and it can be rectified by the Froggatt-Nielsen mechanism~\cite{Froggatt:1978nt}, 
but the introduction of a new symmetry and the breaking of this symmetry at an appropriate 
scale, for this purpose, appear somewhat {\em ad hoc}.  The other issue is that, if one gauges the
U(1)$_{\rm B-L}$ symmetry, three right-handed neutrinos are required to achieve 
the anomaly cancellation.  Then the question arises: what is the role
of the third right-handed neutrino, $N_1$?  In one limiting case, $N_1$
may have a Planck-scale mass, and this particle would play no role in 
the low energy physics. In particular, it is clear from the seesaw
formula that such a heavy $N_1$ does not contribute to the light
neutrino mass, $m_\nu$.  In the other limit, $N_1$ may be very
light. An interesting possibility arises in this case; if the mass is
at a keV scale, then $N_1$, which is long-lived on cosmological time scales, is
a viable dark-matter candidate~\cite{Dodelson:1993je}.  In addition, the same particle would be 
produced in a supernova explosion, and it would be emitted with an anisotropy of a few per cent, which is 
sufficient to explain the observed velocities of pulsars~\cite{Kusenko:1997sp}.
However, one could ask whether such an extremely light mass may upset the 
seesaw mechanism, and it is also unclear how such a small mass may arise 
naturally.  We will show that, in a theory with an extra
dimension, a split mass spectrum arises naturally, without disrupting the
seesaw mechanism.  The {\em split seesaw} mechanism can explain both mild and large mass
hierarchy, i.e., why $M_{2,3} < M_{\rm GUT}$ and why $M_1 \ll M_{2,3}$.

Let us consider a 5D theory compactified on $S^1/Z_2$ with coordinate
$ y \in [0,\ell]$.  One of the boundaries at $y=0$ is identified with
the SM brane, where the SM degrees freedom reside, while the other at
$y=\ell$ is a hidden brane.  The size of the extra dimension $\ell$
and the 5D fundamental scale $M$ are related to the 4D reduced Planck
scale as
\beq
M_P^2 \;=\; M^3 \ell.
\eeq
For the consistency of the theory, the compactification scale $M_c
\equiv 1/\ell$ must be smaller than $M$.  As reference values we take
$M \sim 5 \times \GEV{17}$, $M_c \sim \GEV{16}$ and the $B-L$ breaking
scale $v_{\rm B-L} \sim \GEV{15}$ in the following.

We introduce a Dirac spinor field in 5D, $\Psi = (\chi_\alpha, {\bar
  \psi}^{\dot{\alpha}})^T$, with a bulk mass $m$:
\beq
S\;=\;\int d^4x\, dy \,M \left(i \bar{\Psi} \Gamma^A \partial_A \Psi + m \bar{\Psi}\Psi\right),
\label{5Daction}
\eeq
where $A$ runs over $0, 1,2,3,5$, and the 5D gamma matrices $\Gamma^A$
are defined by
\beq
\Gamma^\mu = \left(
\bear{cc}
0& \sigma^\mu \\
\bar{\sigma}^\mu&0
\eear
\right),~~~~
\Gamma^5 = -i \left(
\bear{cc}
1 &0 \\
0&-1
\eear
\right).
\eeq
We have inserted the mass scale $M$ in Eq.~(\ref{5Daction}) so that
the mass dimension of $\Psi$ is $3/2$ as in the 4D case.  The zero
mode of $\Psi$ should satisfy
\beq
(i\Gamma^5 \partial_5 + m )\Psi^{(0)} \;=\;0.
\eeq
The bulk profile of the zero mode is therefore given by $\exp(\mp m y)$ for $\chi$ and $\bar{\psi}$~\cite{Jackiw:1975fn,Kaplan:1992bt}.

A 4D chiral fermion can be obtained from the 5D Dirac fermion by
orbifolding.  If we assign a $Z_2$ parity $-1$ and $+1$ to $\chi$ and
$\psi$, respectively, we can see that only $\psi$ has a zero mode with
an exponential profile in the bulk. (For consistency we have assigned
a negative $Z_2$ parity to the bulk mass $m$.) The zero mode of $
\Psi_R=(0,\bar{\psi})^T$ can be expressed in terms of the canonically
normalized (right-handed) fermion $\psi_R^{(4D)}$ in the 4D theory as
\beq
\Psi_R^{(0)}(y,x) \;=\; \sqrt{\frac{2m}{e^{2m\ell}-1}} \frac{1}{\sqrt{M}} e^{m y} \psi_R^{(4D)}(x).
\label{relation}
\eeq
As one can see from the $y$-dependence of the wave function, the zero
mode peaks at the hidden brane at $y=\ell$ for positive $m$. In
particular, in the limit of $m \ell \gg 1$, the overlap of the zero
mode with particles on the SM brane becomes exponentially suppressed.

We would like to promote the right-handed neutrino $N_i$ to a bulk
field $\Psi_i$ in 5D~\cite{Hebecker:2002xw}. For definiteness, we identify 
the zero modes of $\Psi_{iR}$ with the right-handed neutrinos in the 4D theory.  
Then the zero modes have an exponential profile in the $y$ direction $\sim
\exp(my)$.  We also introduce a U(1)$_{\rm B-L}$ gauge field in the
bulk and a scalar field $\Phi$ with a $B-L$ charge $-2$ on the SM
brane. Assuming that the $\Phi$ develops a VEV $v_{\rm B-L} \approx
\GEV{15}$, the zero mode of the U(1)$_{\rm B-L}$ gauge boson receives mass via the usual Higgs 
mechanism, and there is no additional zero mode.
The VEV of $\Phi$ also gives rise to heavy Majorana masses for the
right-handed neutrinos.  After integrating out the heavy $\Phi$ and
the U(1)$_{\rm B-L}$ gauge boson, one obtains the Lagrangian for the zero modes:
\bea
S&=& \int d^4x \,dy  \left\{ M\left(i\bar{\Psi}^{(0)}_{iR} \Gamma^A \partial_A \Psi_{iR}^{(0)} + m_i \bar{\Psi}_{iR}^{(0)} \Psi_{iR}^{(0)} \right)
\right.\non\\ 
&& \left.
+ \delta(y) \left(\frac{\kappa_i}{2} v_{\rm B-L}  \bar{\Psi}^{(0)c}_{iR} \Psi^{(0)}_{iR}
 + \tilde{\lambda}_{i\alpha} \bar{\Psi}^{(0)}_{iR}  L_\alpha \,\phi + {\rm h.c.}  \right) \right\} ,
\eea
where $\kappa_i$ and $\tilde{\lambda}_{i \alpha}$ are numerical
constants of order unity, and we introduced the lepton and Higgs
doublets on the SM brane at $y=0$.\footnote{
Since all the right-handed neutrinos are in the bulk, the U(1)$_{\rm B-L}$ 
becomes anomalous on the SM brane. However, one can cancel the anomaly
by introducing Chern-Simons terms with an appropriate coefficient in the bulk~\cite{Izawa:2002qk}.
}  Using the relation
(\ref{relation}), we obtain the effective 4D mass and Yukawa couplings
in (\ref{4dL}) as~\cite{Hebecker:2001wq},
\bea
M_{Ri} &= & \kappa_i v_{\rm B-L} \frac{2m_i}{M(e^{2 m_i \ell}-1)},
\label{5d_mass}
\\
\lambda_{i\alpha} &=& \frac{{\tilde \lambda}_{i \alpha} }{\sqrt{M }} \sqrt{\frac{2 m_i}{e^{2 m_i \ell}-1}}
={\tilde \lambda}_{i \alpha} \sqrt{\frac{M_{Ri}}{\kappa_i v_{\rm B-L}}}.
\label{5d_yukawa}
\eea
As expected, the right-handed neutrino mass and the Yukawa
couplings are suppressed by an exponential factor. In the extreme
limit of $m_i \ell \gg 1$, both $M_R$ and $\lambda$ are extremely
suppressed.  What is remarkable is that the exponential factors 
cancel in the the seesaw formula for the light neutrino masses
(\ref{seesaw}) because the Yukawa coupling squared is suppressed by 
the same factor as the right-handed mass~\cite{Hebecker:2002xw}. Thus, the neutrino masses are given by 
the familiar seesaw relation: 
\beq
\left(m_\nu\right)_{\alpha \beta} \;=\; \left(\sum_i \frac{1}{\kappa_i} \tilde{\lambda}_{i\alpha} \tilde{\lambda}_{i\beta}\right)
 \frac{\la\phi^0\ra^2 }{v_{\rm B-L}},
\label{s-seesaw} 
\eeq
where the quantity in the parenthesis is of order unity. We can
clearly see that no small parameters such as exponentially suppressed
mass or coupling appear in the seesaw formula; the typical neutrino
mass scale is given by the ratio of the square of the weak scale to
the $B-L$ breaking scale, showing that the seesaw mechanism is robust
against splitting the right-handed neutrino mass spectrum.  So we can
easily realize a split mass spectrum by choosing appropriate (not
extremely large or small) values of $m_i$, without spoiling the seesaw
mechanism.  For instance, we obtain $M_{R2} \sim
10^{12}(10^{11})\,{\rm GeV}$ and $M_{R1} \sim$\,keV for $m_{2} 
\simeq 2.3(3.6) \ell^{-1}$ and for $m_1  \simeq 24 \ell^{-1}$, respectively, where we
set $\kappa_i = 1$ and used the reference values of $v_{\rm B-L}$,
$\ell$ and $M$.\footnote{ It is also possible to realize $M_{R,2}
  \simeq M_{R3} \sim {\rm GeV}$ for $m_2 \simeq m_3 \simeq 17
  \ell^{-1}$, as considered in the
  $\nu$MSM~\cite{Asaka:2005an}.
  }

\section{Cosmology}
\label{sec:3}

In the context of sterile neutrino dark matter, it is customary to
parametrize it in terms of the effective mixing angle $\theta$ and the
mass $m_s$. They are given by
\bea
m_s &=& M_{R1},\\
\theta^2&\simeq& \frac{\sum_\alpha |\lambda_{1\alpha}|^2}{M_{R1}^2} \la \phi^0 \ra^2 \non \\
&\simeq& 3 \times 10^{-9} \lrf{\kappa_1^{-1}\sum_\alpha |\tilde{\lambda}_{1 \alpha}|^2}{10^{-4}} \lrf{\GEV{15}}{ v_{\rm B-L}}
\lrf{1{\,\rm keV}}{m_s} , \\
m_\nu &\simeq& \theta^2 m_s, 
\eea
where we have assumed $\theta \ll 1$.  If $N_1$ is the dark-matter particle, the X-ray constraints imply 
$\theta^2 \lesssim 10^{-5} (m_s/{\rm keV})^{-4}$, while the small-scale structure constraints imply $m_s > 2$\,keV
for the resonant production~\cite{Boyarsky:2008mt}  (see, e.g., Refs.~\cite{Kusenko:2009up,Boyarsky:2009ix} for review). For the above range of parameters 
the contribution of $N_1$ to the light neutrino mass is negligible~\cite{Boyarsky:2006jm}.
The prospects of direct detection of sterile dark matter are also stymied by the smallness of the mixing angle~\cite{Ando:2010ye}.
In contrast, the X-ray instruments can search for decays of sterile right-handed neutrinos,  
which have the half-life much longer than the age of the universe.  Here the
smallness of $\sin^2 \theta$ is compensated by the huge number of dark-matter
particles.  In particular, if the 2.5~keV spectral feature in the recent {\em Chandra} X-ray Observatory  
data~\cite{Loewenstein:2009cm} is explained by the decay of a 5~keV relic sterile 
right-handed neutrino,  the inferred parameters $m_s \simeq 5\,$keV and $\theta = (0.2 - 1.4)\times 
10^{-9}$~\cite{Loewenstein:2009cm} imply that such a sterile neutrino does not contribute to the neutrino mass.

To account for the observed dark matter density, sterile right-handed neutrinos must have the correct abundance.  
A generic way to produce relic sterile neutrinos is through non-resonant (NR)
oscillations~\cite{Dodelson:1993je}.  For the mixing angle
$\sin^2 \theta \approx 10^{-9}$ and the mass $m_s \approx 5$\,keV, the 
resulting abundance of sterile neutrinos is in the right range~\cite{Dodelson:1993je,Abazajian:2001nj,Abazajian:2001vt,Asaka:2006nq}. 
Assuming that the sterile neutrinos produced by the NR oscillations
account for the total dark matter density, there is an upper limit $m_s \lesssim 5$\,keV from the X-ray
observations~\cite{Loewenstein:2008yi}.  On the other hand, the lower limits $m_s \gtrsim
8$\,keV~\cite{Boyarsky:2008xj} and $m_s \gtrsim 11$\,keV~\cite{Polisensky:2010rw} based on 
the Ly-$\alpha$ observations and on dwarf spheroidals, respectively, appear to rule out such dark matter produced by NR oscillations. 
Thus, if the X-ray line observed by the {\em Chandra} X-ray Observatory is due to the decay of dark matter sterile neutrinos, they must be
produced by some other mechanism.  Such mechanisms, producing colder sterile neutrinos, consistent with all the constraints, 
have been studied in the context of other models~\cite{Kusenko:2006rh}. 
We will show that, in the split seesaw scenario, the sterile neutrinos can be produced with the same or greater cool-down factor as 
those produced at the electroweak scale~\cite{Kusenko:2006rh}.  Thus, our dark-matter candidate presents no conflict with the 
Lyman-$\alpha$ and other small-scale structure bounds.

In our set-up, the sterile neutrino has a U(1)$_{\rm B-L}$ gauge
interaction, which may become important in the early Universe. In
particular, the reheat temperature $T_R$ must be greater than $M_{R2}$
or $M_{R3}$ for the leptogenesis to work. According to
Ref.~\cite{Raidal:2002xf}, a successful leptogenesis is possible for
$T_R \gtrsim 10^{11}$\,GeV, unless $M_{R2}$ and $M_{R3}$ are extremely
degenerate.  For such a high reheat temperature, the main production
process is pair production of $N_1$ from the SM fermions in
plasma through the s-channel exchange of the B-L gauge boson.  The
production is most efficient at reheating.  The number to entropy
ratio of the sterile neutrino produced by this mechanism can be
roughly estimated as
\bea
Y_{N1}\; \equiv\; \frac{n_{N_1}}{s}  &\sim& \left.\frac{\la \sigma v \ra n_f^2/H}{\frac{2 \pi^2}{45}g_* T^3}\right|_{T=T_R} \non\\
			&\sim & 10^{-4} \lrfp{g_*}{10^2}{\frac{3}{2}} \lrfp{v_{\rm B-L}}{\GEV{15}}{-4} \lrfp{T_R}{5 \times \GEV{13}}{3},
\eea
where $H$ is the Hubble parameter, $g_*$ counts the relativistic
degrees of freedom at the reheating, $\la \sigma v \ra \sim T^2/v_{\rm
  B-L}^4$ is the production cross section, $n_f \sim T^3$ is the
number density of the SM fermions in plasma, and the first equality is
evaluated at the reheating.  The numerical solution of the Boltzmann
equation gives a consistent result~\cite{Khalil:2008kp}.  The dark
matter abundance is related to the mass density to entropy ratio as
\beq
\Omega_{\rm dark}=0.2 \times \left(\frac{m_s}{5\, {\rm keV}}\right) \left( \frac{Y_{N1}}{0.7 \times 10^{-4}} \right).
\label{Omegadm}
\eeq
Thus, the reheat temperature as large as $\GEV{13}$ is needed to
account for all the dark matter density by this production process.
The thermal leptogenesis works with such a high temperature.  Also since the
sterile right-handed neutrinos are out of equilibrium since their production 
at high temperature, when all the Standard model degrees of freedom are thermally excited, 
the average momentum of such dark-matter particles is  red-shifted by a factor of a cubic 
root of the ratio of the degrees of freedom, which is about $3.5$~\cite{Kusenko:2006rh}.

Another scenario for producing dark matter with the correct abundance
(but with different kinetic properties) places no constraint on the
reheat temperature and predicts an even colder dark matter, as long as 
reheating restores the gauge U(1)$_{\rm B-L}$ symmetry.  
We assume that the U(1)$_{\rm B-L}$ symmetry is
broken spontaneously at $10^{15}$~GeV by the VEV of the Higgs boson
$\Phi$.  Since the right-handed neutrinos are coupled to the
$U(1)_{\rm B-L}$ gauge boson with the universal gauge coupling $g_0$, they
are produced at high temperature and they reach the equilibrium density
$n_{\rm eq}(T)$ with a distribution function $f_{\rm eq}(E,T)=
1/(\exp(E/T)+1)$ above the transition temperature $T_t$. However,
after the phase transition, the $U(1)_{\rm B-L}$ gauge boson becomes
massive and the production of $N_i$ is suppressed by the gauge
boson mass (just as in the production mechanism discussed above).
Meanwhile, the rest of the Standard Model particles are produced in
the true vacuum.  The entropy produced in this phase transition can
dilute the number density of dark-matter sterile neutrinos from its
equilibrium value $Y_{N1}^{\rm eq}= 45 \zeta(3) 2/(2 \pi^4 g_*)\sim
10^{-2}$ to what is required to explain dark matter, see
eq. (\ref{Omegadm}). In this case, the reheat temperature is not
constrained from above, but the final temperature after the phase
transition $T_f$ must satisfy the same constraint as the reheat
temperature in the previous scenario.

If the Higgs potential at zero temperature is
\begin{equation}
 V(\phi,0)= -\frac{\mu^2}{2} \phi^2 +\frac{\lambda}{4}\phi^4, 
\end{equation}
where we fix the vev of the Higgs to be $v_{\rm B-L}$, i.e., 
\beq
\frac{\mu}{\sqrt{\lambda}} = v_{\rm B-L} \approx 10^{15}{\rm\,GeV}.
\eeq
The temperature-dependent effective potential~\cite{Kirzhnits:1976ts} is 
\begin{equation}
V(\phi,T)\approx V(\phi,0)+ \frac{3 g^2+4\lambda}{24} T^2 \phi^2 - \frac{3g^3+g \lambda+3 \lambda^{3/2}}{24\pi} T \phi^3, 
\label{potential_T}
\end{equation}
where $g=2 g_0$ is the effective gauge $(B-L)$ coupling of the Higgs boson, which has $(B-L)=-2$.  
The phase transition from the U(1) symmetric vacuum to the broken-symmetry vacuum 
takes place when the tunneling rate~\cite{Kirzhnits:1976ts,Kusenko:1995bw} 
per Hubble volume per Hubble time is of order one: 
\begin{equation}
T_t^4 \lrfp{S_3}{2 \pi T}{\frac{3}{2}} e^{-S_3/T_t} \ \times \ 
H_t^{-4}
\sim 1,
\end{equation}
where 
\beq
H_t \;=\; \lrfp{\pi^2 g_*}{90}{1/2} \frac{T_t^2}{M_P}.
\eeq
Here 
$S_3$ is the action of the three-dimensional bounce, and in the thick-wall limit, 
it is approximately given by~\cite{Kirzhnits:1976ts}
\beq
\frac{S_3}{T} \;\simeq \; \frac{44 \left(\gamma (T^2-T_c^2)\right)^{3/2}}{\alpha^2 T^3},
\eeq
with
\bea
\alpha&\equiv&\frac{3g^3+g\lambda+3\lambda^{3/2}}{8 \pi},~~
\gamma \equiv \frac{3g^2+4\lambda}{12},~~
T_c\equiv\frac{\mu}{\sqrt{\gamma}}.\non
\eea

The correct abundance of dark matter is attained if the ratio of
entropies before and after the phase transition is given by
\begin{equation}
\frac{S_{\rm f}}{S_{\rm t}} \sim \frac{T^3_f}{T^3_t} \sim 10^2, \ \ {\rm or} \ \ T_f\sim 5 T_t.
\label{entropy_production}
\end{equation}
To avoid overproduction of sterile neutrino after the phase
transition, the final temperature should be smaller than the reheat
temperature estimated in the previous scenario,
\begin{equation}
T_f < T_{\rm max} = {5\times10^{13}}\, {\rm GeV} \left(\frac{v_{\rm B-L}}{10^{15} {\rm GeV}} \right)^{4/3}.
\label{T_constraint}
\end{equation}
By energy conservation, $$T_f \sim (\epsilon/g_*)^{1/4}\approx 0.3
\epsilon^{1/4},$$ where $\epsilon=-V(v_{\rm B-L},0)$. Note here that the latent heat comes from
the energy difference between the symmetric phase $\phi=0$ and the broken phase $\phi = v_{\rm B-L}$,
which is estimated by using the zero-temperature potential. 

While there is a significant freedom in the choice of parameters, one can verify that, for example, the
choice of $g\equiv 2 g_0  =0.3$, $\mu=\GEV{11}$ satisfies 
the constraints (\ref{entropy_production},\ref{T_constraint}) for  $T_t\sim 7 \times \GEV{11}$
and $T_f\approx 5 T_t \simeq 3 \times 10^{12}\,$GeV. For an entropy dilution of $O(100)$ to occur,
a relatively light mass $\mu$ is required, which leads to some degree of fine-tuning of the parameters
in the scalar potential. This fine-tuning may be ameliorated by extending the Higgs sector, 
so as to increase the strength of the first-order phase transition. 

While the X-ray constraints depend only on the mass, mixing, and abundance of sterile neutrinos, 
the clustering properties of dark matter, probed by Lyman-$\alpha$ bounds~\cite{Boyarsky:2008xj} 
and the observations of dwarf spheroidal galaxies~\cite{Polisensky:2010rw,Gilmore:2007fy}, are sensitive 
to the momentum distribution of dark-matter particles.  In our first scenario, the limited preheating produces 
a non-thermal distribution of dark matter particles, which is further cooled down 
and redshifted in the subsequent history of the universe.  In our second scenario, in addition to this redshifting, 
the B$-$L breaking phase transition redshifts the momenta of dark-matter particles by 
an additional factor $T_f/T_t\approx 5$.  Thus, the two scenarios produce dark matter 
with different free-streaming lengths, each of which is small enough to satisfy the present bounds.

\section{Discussion}
\label{sec:4}

We have shown that the seesaw formula is robust with respect to splitting the
right-handed neutrino mass spectrum in a model with an extra
dimension. The Yukawa couplings and the Majorana masses are suppressed 
in such a way that the usual seesaw relation is preserved.  This remarkable 
feature allows the use of the seesaw mechanism 
for two heavy right-handed neutrinos and one keV sterile right-handed neutrino 
in a model that does not require any unnatural fine-tuning of parameters. 
The resulting {\em split seesaw} model explains the origin of the baryon asymmetry of the universe, 
dark matter, and the smallness of the active neutrino masses in an natural and holistic manner. 
Furthermore, the light sterile neutrino provides a simple explanation for the observed velocities of pulsars, 
due to anisotropic emission of such particles from a supernova.  Finally, the model offers an explanation as to 
why there are three generations of fermions in nature:  two right-handed neutrinos are
needed for leptogenesis, and the third one plays the role of dark matter.  X-ray astronomy offers 
an exciting opportunity to discover this dark matter candidate, and indeed the recent data from {\em Chandra} 
X-ray Observatory show a spectral feature consistent with this form of dark matter.  If this result is confirmed, the detailed analysis of the 
structure on sub-galactic scales can help distinguish between the two production mechanisms we have discussed.

\begin{acknowledgements}
F.T. thanks Antonio Riotto and CERN Theory Group for the warm hospitality while part of this work was completed.
A.K. was supported  by DOE Grant DE-FG03-91ER40662 and NASA ATP Grant  NNX08AL48G.
The work of F.T. was supported by the Grant-in-Aid for Scientific Research on Innovative Areas (No. 21111006) and 
Scientific Research (A) (No. 22244030), and JSPS Grant-in-Aid for Young Scientists (B)
 (No. 21740160).  This work was supported by World Premier
 International Center Initiative (WPI Program), MEXT, Japan.  A.K. thanks Aspen Center for Physics for hospitality.

\end{acknowledgements}



\end{document}